\documentclass{aa} 
\usepackage{graphicx}
\usepackage{hyperref}
\usepackage{multirow}
\usepackage{makecell}
\usepackage{placeins}
\usepackage{booktabs}

\usepackage{txfonts}

\usepackage{xcolor}
\newcommand{\revcolor}{black}
\newcommand{\rev}[1]{\textcolor{\revcolor}{#1}}

\newcommand{\teff}{$T_\mathrm{eff}$}
\newcommand{\logg}{log\ $g$}
\newcommand{\monh}{$\mathrm{[M/H]}$}

\begin{document} 

   \title{webSME: An online tool to infer stellar parameters and abundances}

   \subtitle{Spectroscopy Made Easy goes online}

   \author{J. Puschnig\inst{1,2}
          \and
          A.J. Korn\inst{2}
          \and
          J. Remmert\inst{2}
          \and
          I. Balkwill-Western\inst{2,3}
          \and
          C. T. Nguyen\inst{4}
          \and
          N. Piskunov\inst{2}
          }

   \institute{
        ICA, Slovak Academy of Sciences, Dubravska cesta 9, 84503 Bratislava, Slovakia
        \and
        Department of Physics and Astronomy, Uppsala University, Box 516, SE-751 20 Uppsala, Sweden
        \and
        Department of Physics, University of Surrey, Guildford, GU2 7XH, Surrey, UK
        \and
        INAF Osservatorio Astronomico di Trieste, Via Giambattista Tiepolo, 11, Trieste, Italy\\
        \email{johannes@jpuschnig.com, andreas.korn@physics.uu.se}
    }

   \date{Accepted July, 2026}

 
  \abstract
   {Stellar spectroscopy is a robust technique for determining fundamental stellar parameters such as effective temperature, surface gravity and metallicity. Spectroscopy Made Easy (SME) has long served as a framework for spectral synthesis and parameter inference. In this paper, we introduce \texttt{webSME}, a web-based extension of the Python implementation of SME. \texttt{webSME} integrates enhancements including a non-local thermodynamic equilibrium abundance correction mode, a precomputed grid of synthetic spectra for robust determination of stellar parameters from large wavelength ranges (thousands of \AA\ wide), Markov Chain Monte Carlo sampling for uncertainty estimation, and support for recent reference abundance patterns. It enables efficient and user-friendly analysis of high-resolution spectra, making it suitable for a wide range of applications - from detailed abundance studies to education and outreach. We demonstrate the performance of \texttt{webSME} on synthetic and observed spectra, including benchmark stars, and validate its accuracy against classical SME-based analysis. The platform's ease of access and advanced capabilities position it as a powerful tool in the modern astrophysical toolkit.}


   \maketitle
%

\section{Introduction}

Stellar spectroscopy is key to unlocking fundamental parameters of stars such as effective temperature (\teff), surface gravity (\logg), and metallicity (\monh). Accurate determination of \teff\ is pivotal because it determines the conditions within the stellar atmosphere and correlates with fundamental physical parameters such as the stellar mass, radius, and luminosity. This, in turn, informs our understanding of stellar evolution, age estimation, and chemical compositions, essential for galactic archaeology \citep{Giribaldi2019}.

Spectroscopic analysis relies on fitting certain spectral lines (or regions). Balmer lines (e.g., H$\alpha$ and H$\beta$) are widely used for \teff\ determination, primarily due to their insensitivity to reddening and minimal dependence on other stellar parameters like \monh\ and \logg\ \citep{Fuhrmann1993,Barklem2000,Barklem2002}. Neutral magnesium (Mg I) is crucial in stellar spectra, especially in late-type stars, where it serves as a tracer of $\alpha$-element abundances \citep{Osorio2015}. The Mg I $b$ triplet (5167--5184 \AA) and the line at 4571 \AA\ are valuable diagnostics for \logg\ and show small deviations from local thermodynamic equilibrium (LTE) in many stars \citep{Carlsson1992,Edvardsson1988}.

Given the complexity of stellar spectra, accurate modeling of radiative transfer is essential. Spectroscopy Made Easy (SME) \citep{Valenti1996, Piskunov2017} offers a comprehensive framework for spectral synthesis and parameter inference using classical model atmospheres \citep{Gustafsson2008}. SME employs user-defined atomic and molecular data (level energies, transition probabilities, broadening parameters) to compute high-resolution synthetic spectra, which can be iteratively fitted to observed spectra to determine stellar parameters and elemental abundances.

This paper presents \texttt{webSME}, an online version of SME designed for improved accessibility and broad application in astrophysics research, including datasets from modern astronomical surveys, and teaching and outreach.

\section{Methods}

The \texttt{webSME} platform utilizes PySME \citep{Wehrhahn2023}, which combines spectroscopic analysis and optimization techniques to provide measurements of key parameters such as effective temperature, surface gravity and metallicity plus auxiliary parameters like micro- and macro-turbulence velocities as well as rotation. Using a least-squares (LSQ) minimization approach, PySME iteratively adjusts synthetic spectra to match observed data by optimizing parameter values to minimize residuals (see section \ref{sec:LSQ} for details). This approach allows PySME to infer both fundamental stellar parameters and elemental abundances with high accuracy. The underlying principles of spectral synthesis in PySME are described in \cite{Wehrhahn2023}.

The \texttt{webSME} platform builds on PySME's robust capabilities by offering an intuitive online interface and server-side computation. It includes several new features aimed at improving both the user experience and the precision of the results:

\begin{itemize}
    \item \textbf{\textit{Gaia}-ESO line lists:} Incorporation of the comprehensive line lists used in the \textit{Gaia}-ESO survey \citep{Heiter2021}. \texttt{webSME} offers the full (atomic plus molecular) \textit{Gaia}-ESO line list as well as lists filtered by recommendation flags and/or lists containing atomic lines only.
    \item \textbf{Asplund 2021 solar reference abundances:} \texttt{webSME} provides the most recent reference solar abundances of \cite{Asplund2021}, allowing to determine relative metallicities based on the most comprehensive and state-of-the-art analysis of the solar photosphere.
    \item \textbf{Precomputed grid mode:} A precomputed grid of synthetic stellar spectra speeds up the analysis of large datasets by providing a first approximation to the stellar parameters, which can then be fine-tuned.
    \item \textbf{MCMC sampling:} Markov Chain Monte Carlo (MCMC) sampling is now available for parameter estimation, improving robustness of the results and providing reliable uncertainties.
\end{itemize}

\texttt{webSME} is developed in Python and integrated into a web-based platform where users can upload normalized spectra and apply parameter inference through an intuitive interface. Its server ensures efficient handling of computationally intensive tasks, although there are inherent limitations due to computational resources.

\subsection{Least Squares Routine}\label{sec:LSQ}
The LSQ routine implemented in PySME (and \texttt{webSME}) differs from standard LSQ due to its unique weighting scheme and optimization algorithm. Standard LSQ typically minimizes the sum of squared residuals with equal or data-dependent weights. However, PySME’s LSQ approach modifies the weights to include residual intensity, prioritizing the continuum over line cores in spectral fitting. This ensures that the optimization does not overly focus on line cores at the expense of the continuum.





{\color{\revcolor}
In the fitting procedure, the normalized residuals $r_i$ and the corresponding ($\chi^2$) statistic between the observed and synthetic spectra are calculated as

\[
r_i =
\frac{\mathrm{obs}_i-\mathrm{synth}_i}
{\mathrm{(S/N)}_i},
\]

\noindent and

\[
\chi^2 =
\sum_{i=1}^{N} \left(
\mathrm{obs}_i\ r_i^2
\right),
\]

where $\mathrm{obs}_i$ and $\mathrm{synth}_i$ denote the observed and synthetic fluxes at wavelength point $i$, respectively, and $(\mathrm{S/N})_i$ represents the corresponding signal-to-noise ratio.
}

Consequently, the residual calculation gives more weight to continuum regions over line cores. This weighting scheme ensures that the fitting process prioritizes a close match in the continuum, while discrepancies in line cores, where classical models perform worse and residuals are relatively larger, have a reduced influence on the overall fit quality.

\subsection{LSQ Uncertainties}
The uncertainties of the fit parameters are derived using a combination of numerical, measurement, and model error estimates. Initially, formal uncertainties are calculated from the covariance matrix using the Levenberg-Marquardt algorithm, which assumes that measurement errors dominate. However, these formal uncertainties tend to underestimate the true error when model errors are significant. To address this, SME also provides a heuristic uncertainty estimate that accounts for model errors by identifying pixels sensitive to parameter changes. Pixels are considered sensitive if they meet two criteria: (1) their partial derivative with respect to the parameter being evaluated is non-zero, meaning changes in the parameter affect the flux at that pixel, and (2) the residual between observed and model fluxes is less than five times the measurement uncertainty, ensuring that the sensitivity is not dominated by noise. For each sensitive pixel, SME computes the parameter change needed to match the observed flux, assuming a linear response. These individual parameter shifts are then used to construct a cumulative distribution, providing a robust measure of uncertainty that accounts for model errors such as incomplete line data or simplified physics. This approach offers a more realistic estimate of the true uncertainty compared to purely formal methods.

\subsection{Precomputed Grid Mode}
{\color{\revcolor}
The precomputed grid of synthetic spectra currently distributed with \texttt{webSME} employs MARCS model atmospheres \citep{Gustafsson2008}. It covers the parameter ranges described below, namely $T_{\rm eff}=3600$--$7600$~K, $\log g=1.0$--$5.0$, and $\mathrm{[M/H]}=-3.5$ to $+0.5$.
Alternative model atmospheres such as PHOENIX \citep{Hauschildt2025} are not part of the present precomputed grid; but they could in principle be used in the future.
}

To enable efficient inference of stellar parameters from large, high-resolution spectra spanning e.g.\ more than 1000\,\AA, we generated an extensive grid of synthetic spectra. This approach allows for rapid parameter extraction from broad spectral coverage, a process that would typically require several days of computation but can now be completed in approximately 10 hours on our server. The synthetic spectra are calculated using non-LTE models and the \textit{Gaia}-ESO atomic line list. The spectral grid spans two primary wavelength ranges: 4750-6850\,\AA\ and 8488-8950\,\AA, as defined by the \textit{Gaia}-ESO line list. This vast dataset, stored in a 600GB HDF5 file, encompasses a comprehensive parameter space for surface gravity (\(\log g\)) from 1.0 to 5.0 in steps of 0.5, effective temperature (\(T_{\text{eff}}\)) from 3600 to 7600 K in increments of 250 K, metallicity (\([\text{M/H}]\)) from $-$3.5 to +0.5 in 0.5 steps, macroturbulence (\(v_{\text{mac}}\)) from 0.0 to 15.0 km/s in 3.0 km/s increments, and microturbulence (\(v_{\text{mic}}\)) from 0.0 to 3.0 km/s with a step size of 1.0 km/s. This expansive, structured grid allows for rapid spectral fitting and parameter inference, significantly enhancing the analysis of stellar spectra with broad wavelength coverage.

In precomputed grid mode, after identifying the initial best-fit point, a refined subgrid is constructed around it to adjust all five stellar parameters with high precision. Using linear interpolation across these five dimensions - \(T_{\text{eff}}\), \(\log g\),\([\text{M/H}]\),\,\(v_{\text{mac}}\) and \(v_{\text{mic}}\) - the subgrid allows for smooth transitions between neighboring grid points. The refinement introduces a higher resolution for each parameter:

\begin{itemize}
    \item \( T_{\text{eff}} \): varies within \(\pm 125\) K in steps of 75 K,
    \item \( \log g \): varies within \(\pm 0.25\) in increments of 0.15,
    \item \([\text{M/H}]\): varies within \(\pm 0.25\) in increments of 0.15,
    \item \( v_{\text{mac}} \): varies within \(\pm 1.5\) km/s in steps of 1.0 km/s,
    \item \( v_{\text{mic}} \): varies within \(\pm 0.5\) km/s in increments of 0.3 km/s.
\end{itemize}

When inferring all five parameters, the subgrid comprises 1024 points, each requiring a linear interpolation. As tests have shown, this setup provides sufficiently detailed sampling around the initial best-fit. It is important to note that uncertainties are not calculated in the precomputed grid LSQ  mode. Accurate uncertainties are available through MCMC sampling.

\subsection{Markov Chain Monte Carlo}
We use the above precomputed grid for the MCMC part, as synthesizing the spectra along the chain would incur significant additional computational time. Our parameter inference approach within the MCMC framework relies on three main components: the likelihood, the prior, and the posterior probability. Each of these is defined through specific functions to evaluate parameter values (\(\theta\)) against observed data, while ensuring that parameter boundaries are respected.

\begin{itemize}
    \item \textbf{Log-Likelihood Function:} The log-likelihood function quantifies how well the model with up to five parameters \(\theta\) matches the observed data. This is achieved by calculating residuals between the synthetic and observed spectra, then computing the chi-squared statistic, \(\chi^2 = \sum \texttt{residuals}^2\). The function returns \(-0.5 \chi^2\) as the log-likelihood, which becomes a large negative value if \(\chi^2\) is non-finite, effectively rejecting the parameter set.

    \item \textbf{Log-Prior Function:} The log-prior function encodes our prior knowledge or constraints on parameter values. Each parameter \(\theta_i\) is checked against specified bounds, defined by our precomputed grid. If any parameter value falls outside these bounds, the function returns \(-\infty\), making the parameter set unlikely. Otherwise, it returns 0, indicating a uniform prior within the allowed range.

    \item \textbf{Log-Posterior Function:} The log-posterior combines the log-prior and log-likelihood to calculate the log-posterior probability, which is the quantity maximized during sampling. It first evaluates the log-prior for \(\theta\); if the prior is non-finite (i.e., outside bounds), it returns a large negative value to reject the sample. If the prior is acceptable, it proceeds to calculate the log-likelihood. If both the prior and likelihood are finite, their sum is returned as the log-posterior probability. This value guides the MCMC sampler by weighting parameter sets according to their probability, steering the search toward the most likely regions in parameter space.
\end{itemize}

\bigskip

In practice, we encountered a convergence issue in which the MCMC chains struggled to efficiently explore the parameter space, often taking a long time to reach high-probability regions. This was likely due to the initial positions of the walkers being too far from the optimal areas in parameter space, causing the chains to converge slowly. 
To address this, we implemented a staged initialization strategy:

\begin{enumerate}
    \item Run a short preliminary chain of 50-200 iterations to allow the walkers to explore the parameter space.
    \item Identify the position of highest probability from this short run and reinitialize the walkers in a small elliptical region around this point.
    \item Optionally, repeat this step to ensure the walkers are sufficiently close to the high-probability region.
    \item Finally, run the full MCMC chain from these optimized starting points.
\end{enumerate}

This approach accelerates convergence by guiding the walkers toward regions with high probability before beginning the full sampling process, thereby improving the efficiency of the parameter inference.

\bigskip

To determine when the chain has converged and can terminate, we set a criterion based on the autocorrelation time, \(\tau\). The chain stops if either (1) the chain length reaches 50 times the autocorrelation time, \(50 \times \tau\), indicating that sufficient independent samples have been gathered, or (2) a maximum of 1000 steps is completed, ensuring termination in cases where convergence might be slower. This termination criterion ensures a balance between accuracy and computational efficiency. Additionally, the chain can be resumed at any time from where it stopped, allowing for flexible continuation of the sampling process if further refinement is required.

\bigskip

\subsection{MCMC Uncertainties}

\begin{figure*}
    \centering
    \includegraphics[width=\textwidth]{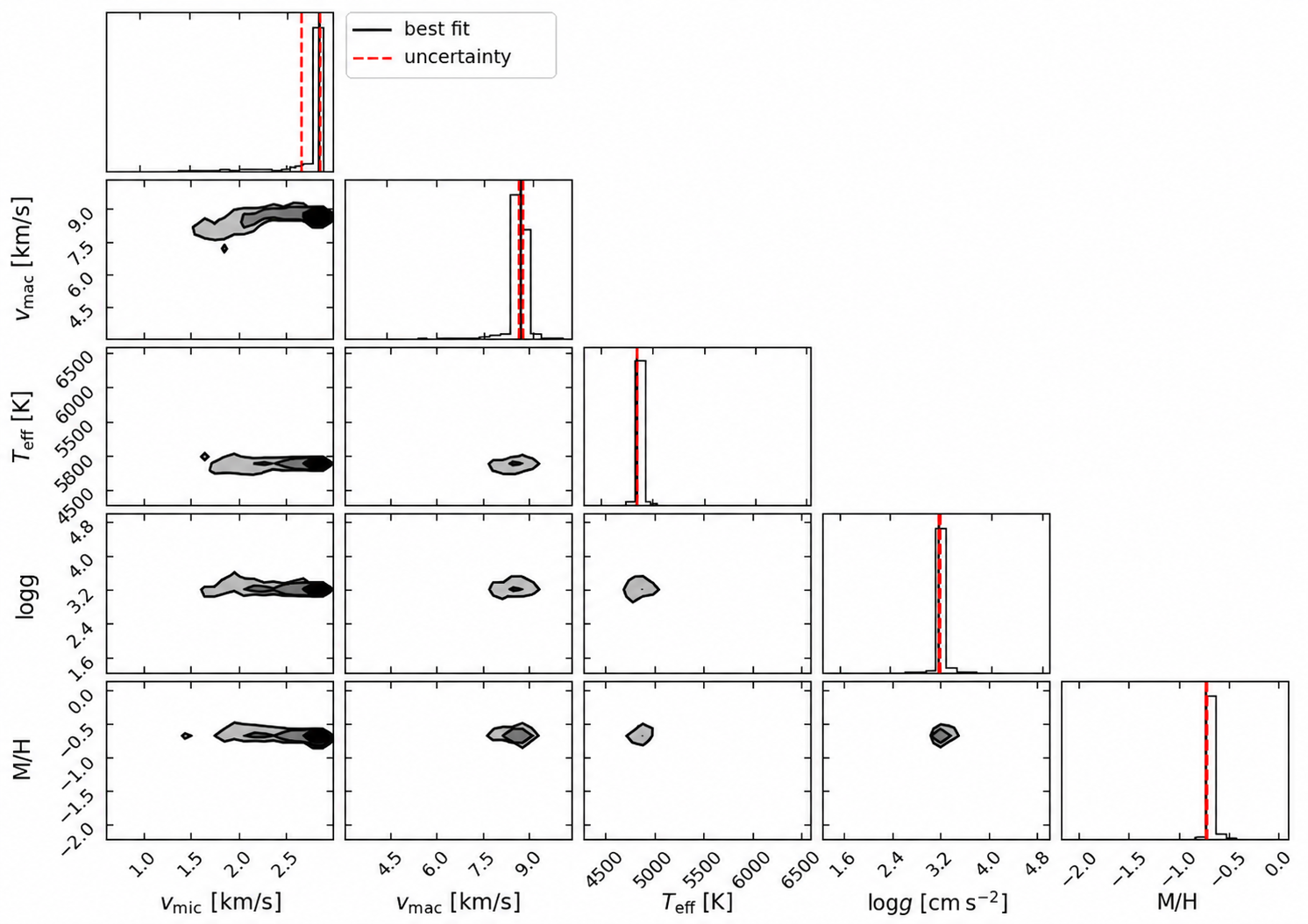}
    {\color{\revcolor}
    \caption{Stellar parameter inference using Monte Carlo Markov Chain sampling applied to a synthetic input spectrum with known stellar parameters: \( T_{\text{eff}} \) = 4856 K, \( \log g \) = 3.2 cm s$^{-2}$, \([\text{M/H}]\) = -0.7 and \( v_{\text{mic}} \) = 2.7 km s$^{-1}$. Furthermore, noise was added to the synthetic spectrum to mimique observations with a signal-to-noise ratio of 100.}
    }
    \label{fig:mcmc}
\end{figure*}

Once the MCMC sampling is complete, uncertainties for each parameter are estimated by analyzing the distribution of sampled values. For each parameter, a Gaussian kernel density estimate (KDE) is applied to the posterior samples to smooth out the distribution and locate the most probable value. The mode of the KDE distribution represents the best-fit parameter value. To quantify uncertainties, we compute the 16th, 50th, and 84th percentiles of the parameter samples. The 50th percentile is the median value, while the difference between the median and the 16th and 84th percentiles provides the lower and upper bounds of the 1\(\sigma\) credible interval. Figure \ref{fig:mcmc} shows that this approach offers a robust measure of central tendency and credible intervals for each parameter, reflecting the confidence in the inferred values.

\subsection{Spectrum Normalization}

For accurate analysis, \texttt{webSME} requires a normalized spectrum as input, ideally accompanied by a mask that specifies pixel designations. The mask must be of integer type and uses the following pixel codes:

\begin{itemize}
    \item \textbf{0}: Bad pixel
    \item \textbf{1}: Good pixel or line pixel (when the mask also designates continuum pixels)
    \item \textbf{2}: Continuum pixel
\end{itemize}

The interpretation of the mask depends on its values. If the mask contains only \([0,1]\), it serves solely to distinguish between bad pixels \([0]\) and good pixels \([1]\). However, if the mask includes values \([0,1,2]\), it provides a more refined designation, distinguishing bad pixels \([0]\), line pixels \([1]\), and continuum pixels \([2]\). This differentiation is crucial for fitting and parameter inference, as it ensures that appropriate parts of the spectrum are used for continuum placement and feature analysis.

To facilitate the preparation of spectra for \texttt{webSME}, the \texttt{Normalizer} tool is recommended, available at \url{https://github.com/astrojohannes/normalizer}. The \texttt{Normalizer} efficiently processes spectra, ensuring correct flagging and data formatting according to \texttt{webSME} requirements, making it the preferred tool for spectrum normalization in this context.

{\color{\revcolor}
\subsection{Rotation and line broadening.}
In \texttt{webSME}, stellar rotation is modeled via a rotational broadening kernel parameterized by $v\sin i$, combined with instrumental and, optionally, macroturbulent broadening. Users may either supply $v\sin i$ as a fixed input or include it among the parameters to be optimized. In the analyses presented here, $v\sin i$ is not the parameter of interest and is kept fixed during the line-by-line abundance determinations, because simultaneous fitting of rotation and macroturbulence becomes degenerate at the spectral resolutions and \rev{S/N} ratios used in stellar spectroscopy.
}

\subsection{Typical server runtime}
\label{sec:runtime}
The runtime of \texttt{webSME} varies significantly, with differences spanning from minutes to several days. This wide range is heavily influenced by the choice of line list and wavelength coverage. Fastest results are enabled by using the \textit{Gaia}-ESO (Y,Y|U), atomic line list which is the \textit{Gaia}-ESO atomic line list filtered by the recommendation flags as defined in \citet{Heiter2021}. The \textit{Gaia}-ESO (Y,Y|U), atomic line list selection contains only lines with a recommendation flag (Y) for the gf-value and a recommendation or undecided flag (Y|U) for blending. In this configuration, \texttt{webSME} completes inference of up to six free parameters from even very large wavelength ranges of up to 2000~\AA\ within less than an hour. However, the number of synthesized lines is limited ($\sim$1,200 lines across 2000~\AA), which may be insufficient, depending on the science case.

Using the full atomic \textit{Gaia}-ESO line list leads to a significant increase in the number of synthesized lines. Typically $\sim$33,000 lines are found within ranges of 1000~\AA. Tasks based on the \textit{Gaia}-ESO atomic line list and a wavelength range of $\sim$1000~\AA\ may thus run up to $\sim$120 hours. Note that the precomputed grid was calculated using the full atomic \textit{Gaia}-ESO line list and thus may be used for the inference of up to five stellar parameters based on very large wavelength ranges of up to 2000~\AA. Almost independent on the wavelength range, the runtime in the precomputed grid mode is approximately 10 hours.

Including molecular lines by selection of the complete \textit{Gaia}-ESO line list, dramatically increases line density, with typically 130,000 lines per 20~\AA. Inference of up to six parameters from a range of 20~\AA\ may thus take 24 hours when the complete \textit{Gaia}-ESO line list is selected.

\subsection{Post-Processing}
We developed a tool that enables predictions from theoretical stellar models covering a wide metallicity range from $Z = 0.0001$ to $0.02$ and stellar masses from $0.64\,M_\odot$ to approximately $2\,M_\odot$. The models incorporate key internal mixing processes - such as atomic diffusion, first dredge-up, thermohaline mixing, and rotation - that significantly influence surface abundance variations throughout stellar evolution. The stellar evolutionary tracks are based on the PARSEC database\footnote{\url{https://stev.oapd.inaf.it/PARSEC/}}$^{,}$\footnote{\url{https://zenodo.org/records/13283724}}, with detailed descriptions provided in \citet{2012MNRAS.427..127B,2022A&A...665A.126N,2024arXiv240805039N}.

The tool, as implemented in \texttt{webSME}, enables an estimation of the evolution of individual chemical elements. Additionally, it provides a prediction of the star’s current evolutionary stage and ongoing mixing processes via Kiel diagram fitting. Further details on the fitting methodology and the chemical elements included in the dataset are available in the documentation accessible through the \texttt{webSME} interface.

\section{Testing stellar parameter inference}\label{sec:testing_inference}
This section describes the capabilities of \texttt{webSME} to infer stellar parameters from input spectra with known stellar parameters. In the first part, we utilize previously synthesized spectra and study the robustness of \texttt{webSME} against an increasing noise level. In the second part, VLT/UVES spectra of two \textit{Gaia} FGK benchmark stars, HD84937 and $\beta$ Gem, are used as input \citep{Blanco-Cuaresma2014}.

\subsection{Robustness against noise}\label{sec:testing_inference_noise}
In the following, we describe the tests we performed to evaluate \texttt{webSME}'s ability to infer stellar parameters from increasingly noisy data. To do so, we use both the pySME LSQ routine (i.e. scipy.optimize.least\_squares) and our precomputed grid LSQ approach. The test spectra were first generated using \texttt{webSME}'s forward modeling mode, employing the \textit{Gaia}-ESO line list, which is the exact line list used for the creation of the precomputed grid. After generating the synthetic spectra, they were degraded using Gaussian noise corresponding to signal-to-noise ratios of 100, 50, and 25 to simulate realistic observation conditions. The tests were further conducted for two sets of stellar parameters: one on-grid (Test 1) and one off-grid (Test 2), allowing a comparison of the performance of the standard least-squares fitting and the precomputed grid fitting methods. The results are summarized in Tables \ref{tab:noisetest1} and \ref{tab:noisetest2} and show that the least-squares method can accurately infer parameters at high \rev{S/N} but degrades significantly as noise increases, while the precomputed grid method maintains robust performance across all noise levels, although limited by the grid resolution.

\renewcommand{\arraystretch}{1.25} 

\begin{table*}[h!]
\caption{Testing stellar parameter derivation as a function of \rev{S/N}, using  different solving routines: the pySME solver based on \texttt{scipy.optimize.least\_squares} (top rows) and \texttt{webSME}'s LSQ routine on the precomputed grid (bottom rows). In this case, the input parameters are matching one of the grid points.}
\label{tab:noisetest1}
\centering
\begin{tabular}{|c|c|c|c|c|c|c|}
\hline
\textbf{4800-5800\AA} & \textbf{\rev{S/N}} & \textbf{Teff (K)} & \textbf{logg} & \textbf{[M/H]} & \textbf{vmic (km/s)} & \textbf{vmac (km/s)} \\ \hline
\multicolumn{2}{|c|}{\textbf{True Parameters}} & \textbf{5600} & \textbf{4.5} & \textbf{-2.0} & \textbf{1.0} & \textbf{9.0} \\ \hline
\multirow{4}{*}{on-grid, pySME solver} & inf & 5553.31 ± 48.13 & 4.53 ± 0.10 & -1.99 ± 0.04 & 0.00 ± 0.01 & 8.23 ± 0.56 \\ \cline{2-7} 
 & 100 & 5628.25 ± 159.99 & 4.51 ± 0.46 & -1.93 ± 0.17 & 0.93 ± 0.62 & 8.78 ± 2.59 \\ \cline{2-7} 
 & 50 & 5739.22 ± 377.70 & 4.51 ± 1.07 & -1.71 ± 0.41 & 0.86 ± 1.46 & 10.06 ± 6.49 \\ \cline{2-7} 
 & 25 & 6072.53 ± 1214.82 & 3.78 ± 4.31 & -0.75 ± 1.04 & 1.01 ± 2.50 & 15.95 ± 20.86 \\ \hline

\multirow{4}{*}{\makecell{on-grid, webSME solver \\ on precomputed grid \\ nearest neighbor}} & inf & 5600.00 & 4.50 & -2.00 & 1.00 & 9.00 \\ \cline{2-7} 
 & 100 & 5600.00 & 4.50 & -2.00 & 1.00 & 9.00 \\ \cline{2-7} 
 & 50 & 5600.00 & 4.50 & -2.00 & 1.00 & 9.00 \\ \cline{2-7} 
 & 25 & 5600.00 & 4.50 & -2.00 & 1.00 & 9.00 \\ \hline

\end{tabular}
\end{table*}

\begin{table*}[h!]
\caption{Testing stellar parameter derivation as a function of \rev{S/N}. In this case the input spectrum was synthesized to not match any grid point.}
\label{tab:noisetest2}
\centering
\begin{tabular}{|c|c|c|c|c|c|c|}
\hline
\textbf{4800-5300\AA} & \textbf{\rev{S/N}} & \textbf{Teff (K)} & \textbf{logg} & \textbf{[M/H]} & \textbf{vmic (km/s)} & \textbf{vmac (km/s)} \\ \hline
\multicolumn{2}{|c|}{\textbf{True Parameters}} & \textbf{4856} & \textbf{3.2} & \textbf{-0.7} & \textbf{2.7} & \textbf{8.5} \\ \hline
\multirow{4}{*}{\makecell{off-grid, webSME solver \\ on precomputed grid \\ nearest neighbor}} & inf & 5100.00 & 3.50 & -0.50 & 3.00 & 9.00 \\ \cline{2-7} 
 & 100 & 5100.00 & 3.50 & -0.50 & 3.00 & 9.00 \\ \cline{2-7} 
 & 50 & 5100.00 & 3.50 & -0.50 & 3.00 & 9.00 \\ \cline{2-7} 
 & 25 & 5100.00 & 3.50 & -0.50 & 3.00 & 9.00 \\ \hline
\end{tabular}
\end{table*}

\subsection{Benchmark stars}\label{sec:testing_benchmark}

We performed several tests with \texttt{webSME} to evaluate its ability to infer stellar parameters from large wavelength ranges, finding that flagging - particularly of telluric (or bad, noisy) regions - is crucial for achieving accurate results. Proper flagging enables the fitting algorithm to ignore noisy or unreliable parts of the spectrum, leading to more reliable parameter estimations.

We utilize the precomputed grid in these tests, as we are investigating high-resolution spectra over large wavelength ranges. Note that the grid spectra have a fixed spectral resolution of $R$\,=\,100,000. As a result, the inferred macroturbulence may not only represent large-scale broadening by stellar surface convection, as the UVES spectra of HD84937 and $\beta$ Gem have a lower/higher resolution of R=74,450R/107,200 respectively, affecting the direct comparability of line broadening effects between the observed and model spectra.

\subsubsection{HD84937}

For the metal-poor benchmark star HD84937, we conducted several specific tests:

\begin{itemize}
    \item \textbf{Test 1}: In this test, the wavelength range was limited to 5840-6620\,\AA, and no flagging was applied. The inferred parameters showed reasonable agreement with the benchmark values, but there were some deviations, particularly in surface gravity and microturbulence.

    \item \textbf{Test 2}: For this test, we used the same wavelength range of 5840-6620\,\AA\ as in Test 1, but applied flagging for tellurics, continuum, and line regions. This setup produced improved results for effective temperature but showed deviations in surface gravity and metallicity.

    \item \textbf{Test 3}: Here, the wavelength range was widened to
    {\color{\revcolor}
    4800-5800\,\AA,
    }
    with flagging to identify tellurics, continuum, and line regions. This setup provided accurate results across all parameters, with inferred values for effective temperature, surface gravity, metallicity, and microturbulence closely matching the benchmark values.
    A small part of the spectrum is shown in Figure \ref{fig:hd84937_betgem_obs_vs_synth}.

    \item \textbf{Test 4}: In this case, the spectral range was expanded to 4800-6800\,\AA, using the same flagging as in Test 3. Although the results remained reasonably accurate, the broader wavelength range introduced greater variability in parameter accuracy.
\end{itemize}

Table~\ref{tab:benchmark1} summarizes the inferred parameters from each test, compared to the benchmark values for HD84937. Overall, Test 3 provided results closest to the benchmark values, demonstrating that a focused wavelength range with appropriate flagging allows \texttt{webSME} to yield accurate stellar parameters. By contrast, Test 4 showed that an expanded range, while informative, does not necessarily lead to improved accuracy.

\begin{table}[h!]
\centering
{\color{\revcolor}
\caption{Comparison of inferred parameters with benchmark values for HD~84937. Since grid mode was used in these cases, the uncertainties correspond to half the grid spacing, i.e. \(\pm 38\,\mathrm{K}\) for \(T_{\mathrm{eff}}\), \(\pm 0.075\,\mathrm{dex}\) for \(\log g\) and \([\mathrm{M/H}]\), and \(\pm 0.15\,\mathrm{km\,s^{-1}}\) for \(v_{\mathrm{mic}}\).}
}
\label{tab:benchmark1}
\[
\begin{array}{|c|c|c|c|c|c|}
\hline
\text{Parameter} & \text{Benchmark} & \text{T1} & \text{T2} & \text{T3} & \text{T4} \\
\hline
T_{\text{eff}} & 6356 & 6225 & 6375 & 6300 & 6300 \\
\log g & 4.06 & 2.25 & 2.55 & 3.75 & 3.55 \\
\text{[M/H]} & -2.09 & -2.10 & -1.60 & -2.10 & -1.95 \\
v_{\text{mic}} & 1.29 & 0.00 & 0.00 & 1.40 & 0.80 \\
\hline
\end{array}
\]
\end{table}

\subsubsection{$\beta$ Gem}

We performed several tests with \texttt{webSME} on the benchmark star $\beta$ Gem, using the precomputed grid.

\begin{figure*}
    \centering
    \includegraphics[width=\textwidth]{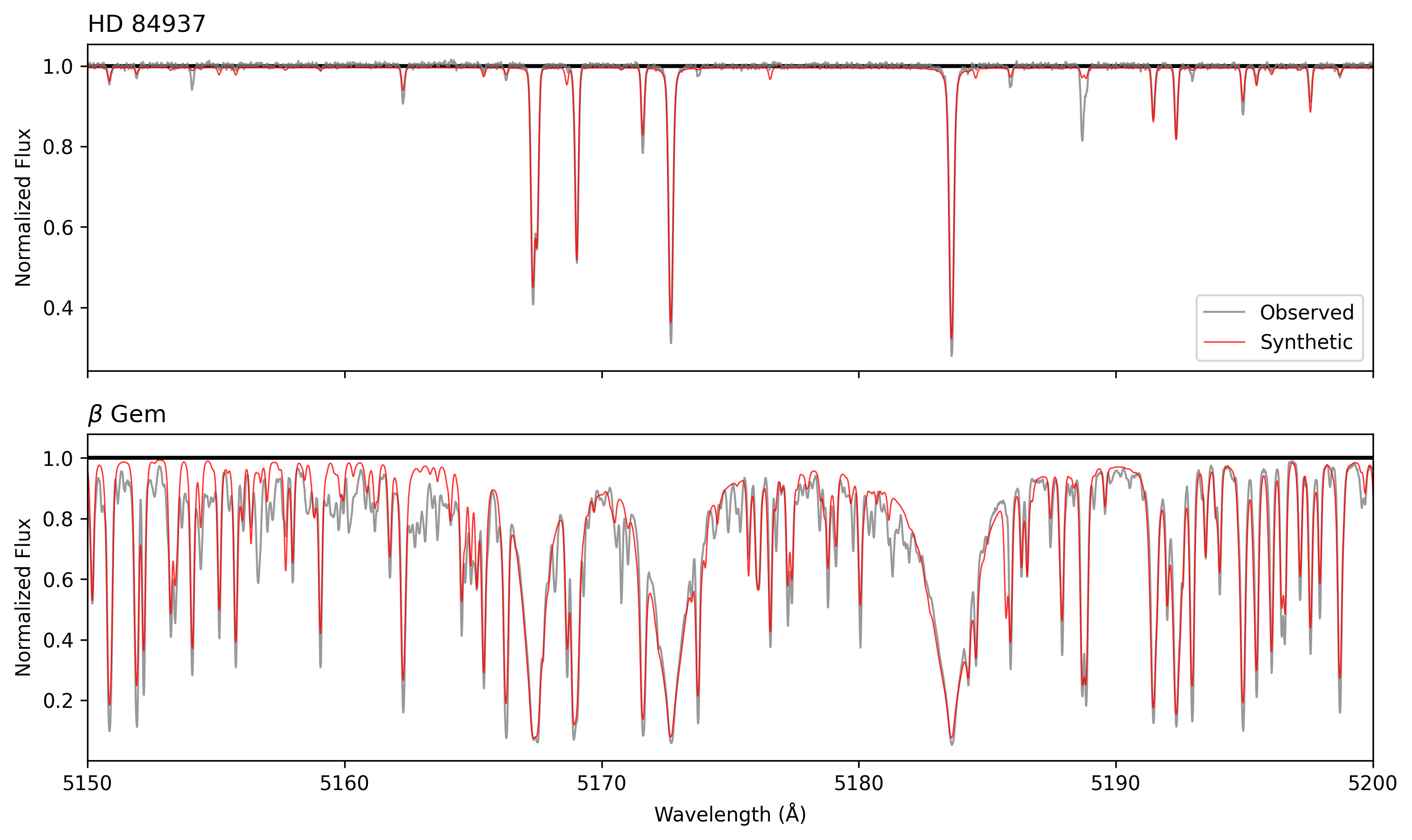}
    {\color{\revcolor}
    \caption{Observed spectral range (gray) versus best-fit result from \texttt{webSME} (red) for HD84936 (top) and $\beta$ Gem (bottom). Shown are cutouts around the Mg I $\lambda$ 5167-5183 magnesium triplet.}
    \label{fig:hd84937_betgem_obs_vs_synth}
    }
\end{figure*}

\begin{itemize}
    \item \textbf{Test 1}: We used a wavelength range of 4800-6800\,\AA\ as input for \texttt{webSME}, without applying any flagging. The inferred parameters showed reasonable agreement with the benchmark values, though all parameters exhibited minor deviations.

    \item \textbf{Test 2}: Using the same wavelength range of 4800-6800\,\AA, we applied flagging for bad, line, and continuum regions. This setup provided the best match across all parameters, with very close values for \( T_{\text{eff}} \), \( \log g \), and minor deviations in [M/H] and \( v_{\text{mic}} \). A small part of the spectrum is shown in Figure \ref{fig:hd84937_betgem_obs_vs_synth}.

    \item \textbf{Test 3}: For this test, we restricted the wavelength range to 4800-5800\,\AA. The inferred parameters show some discrepancies, particularly with lower values for effective temperature and metallicity.

    \item \textbf{Test 4}: Here, we used the wavelength range of 5800-6800\,\AA. The results show a good match for \( T_{\text{eff}} \) and \( \log g \), while metallicity ([M/H]) was slightly underestimated, and \( v_{\text{mic}} \) was slightly overestimated.
\end{itemize}

Table~\ref{tab:benchmark2} summarizes the inferred parameters from each test compared to the benchmark values for $\beta$ Gem. Test\,2 yields the most accurate results, particularly when flagging is applied over the 4800-6800\,\AA\ wavelength range. Flagging enhances accuracy by ensuring that \texttt{webSME} focuses on clean spectral regions and avoids areas affected by noise or artifacts. This test demonstrates that careful wavelength selection and proper flagging can significantly improve the accuracy of inferred stellar parameters when using \texttt{webSME}.

\begin{table}[h!]
\centering
{\color{\revcolor}
\caption{Comparison of inferred parameters with benchmark values for $\beta$ Gem. Since grid mode was used in these cases, the uncertainties correspond to half the grid spacing, i.e. \(\pm 38\,\mathrm{K}\) for \(T_{\mathrm{eff}}\), \(\pm 0.075\,\mathrm{dex}\) for \(\log g\) and \([\mathrm{M/H}]\), and \(\pm 0.15\,\mathrm{km\,s^{-1}}\) for \(v_{\mathrm{mic}}\).}
}
\label{tab:benchmark2}
\[
\begin{array}{|c|c|c|c|c|c|}
\hline
\text{Parameter} & \text{Benchmark} & \text{T1} & \text{T2}  & \text{T3} & \text{T4} \\
\hline
T_{\text{eff}} & 4858 & 4975 & 4800 & 4700 & 4800 \\
\log g & 2.90 & 2.75 & 2.90 & 2.70 & 2.90 \\
\text{[M/H]} & 0.12 & 0.25 & 0.05 & -0.10 & -0.10 \\
v_{\text{mic}} & 1.22 & 1.10 & 1.10 & 1.40 & 1.40 \\
\hline
\end{array}
\]
\end{table}

\section{Element abundance inference}
\label{sec:abund}

In \texttt{webSME}, users have the choice to select individual elements for abundance inference. When an element is chosen, its abundance is treated as a free parameter during the minimization process. This means that the abundance value of the selected element is adjusted iteratively -- alongside other stellar parameters if necessary -- in order to minimize the residuals between the observed and synthetic spectra. By selecting or limiting the spectral range to a region that is most sensitive to the chosen element, \texttt{webSME} is able to precisely determine an element abundance while holding other elements fixed or scaled according to a reference pattern. This targeted approach (e.g. choosing a single-line) is particularly useful when specific chemical tracers are of interest, or when the global metallicity does not adequately reflect the individual abundance pattern of the star.

{\color{\revcolor}
We performed extensive consistency checks between \texttt{webSME} -- which is based on pySME 1.0.1 as of July 2026 -- and the latest version of SME involving regular users/developers of (py)SME in Uppsala, Stockholm and Malmö. In particular, abundance-determination tests demonstrate that the results obtained with \texttt{webSME} and SME are consistent, with inevitable numerical differences that remain well below the reported statistical uncertainties.
}

Note that \texttt{webSME}'s default starting abundances follow the \cite{Asplund2021} solar abundance scale. However, users can modify the initial abundances (in the "H=12" scale), which can improve the efficiency of the inference process by starting from values closer to the expected solution.

\subsection{Non-LTE Abundance Correction Mode}\label{sec: nlte abundance corr mode}
\texttt{WebSME} not only enables the inference of elemental abundances by fitting model parameters to observed spectra, but also includes a non-LTE (NLTE) abundance correction mode. Spectral lines synthesized under the assumption of LTE are often in systematic error \citep[e.g.][]{2024ARA&A..62..475L}. These errors depend sensitively on the stellar parameters, while their size (and even their sign) can be quite different for different lines of the same atomic or ionic species.  For example, the cores of the \ion{O}{~I} 777 nm triplet lines are usually too weak in LTE, as photon losses lead to an overpopulation of the metastable lower level and extra line opacity \citep[e.g.][]{Amarsi2016a}, leading to negative abundance corrections. Whereas the \ion{Cu}{~I} 327.4~nm resonance line is sensitive to over-ionisation of the minority neutral species and the line is too strong in LTE, leading to positive abundance corrections \citep[e.g.][]{Caliskan_2025}. Therefore, it is crucial to consider the LTE/NLTE assumptions made when interpreting spectral features, as the resulting abundance estimates may differ significantly.\par
The NLTE abundance correction mode computes the difference in elemental abundance estimates derived from LTE and NLTE models for a given element. This is done by determining the difference between LTE and NLTE abundances through matching of the equivalent width (EW) of the NLTE synthetic line to that of the LTE line. To estimate the NLTE correction, the user specifies the $T\mathrm{_{eff}}$, log $g$, $\mathrm{[M/H]}$ and $v_\mathrm{mic}$ to be used. Additionally, the line list, wavelength range and integration range, as well as the specific elemental composition to be used in the synthesis is set by the user. This also allows the user to tailor the abundance of the element under consideration ($\mathrm{X}$) independently of the set $\mathrm{[M/H]}$, as the NLTE effects typically are a strong function of $\mathrm{[X/Fe]}$.\newline
Finally, the method fixes the LTE EW based on the set input parameters and iteratively adjusts the abundance in the NLTE model until the corresponding EW matches the LTE value. This process begins with a linear extrapolation of the curve of growth (COG) to obtain an initial estimate of the target abundance. Then, several data points along the COG are computed, allowing the use of cubic spline interpolation. The final target abundance is subsequently determined by finding the root of the function defined as $f(a)=\mathrm{COG}(a)-\mathrm{EW_{LTE}}$.\par
Currently, \texttt{WebSME} and PySME utilize precomputed grids of departure coefficients from \cite{Amarsi_2020,Amarsi_2022}, \cite{Mallinson_2024}, and \cite{Caliskan_2025}. These grids of departure coefficients were calculated with the NLTE code \texttt{Balder} \cite{Amarsi_2018} using the MARCS atmospheric model \cite{2008A&A...486..951G}. The model atoms, described in those papers, all employ modern descriptions for inelastic collisions with neutral hydrogen based on full quantum-mechanical models, or on asymptotic models often in combination with the free electron model \citep[see e.g.][for a recent discussion and see Table~\ref{tab:NLTE atoms + grids} for a summary of model atoms and grids employed by \texttt{webSME}]{2024PhRvA.109e2820S}.\par
This approach enables computation times that are typically under a minute, often between 20 and 30 seconds. Other publicly available tools also compute NLTE abundance corrections, but are usually limited to a fixed set of spectral lines and abundance that scale with the atmospheric metallicity \citep[e.g. MPIA-NLTE;][]{Kovalev2019}. In contrast, \texttt{webSME} supports user-defined line lists, can handle blended lines through optional integration range inputs, and gives users full control over stellar input abundances. Custom line lists must be provided in VALD3 format \citep{2015PhyS...90e4005R}. The lists have to be in so-called long format, which includes the term designations necessary for the NLTE calculations.  {\color{\revcolor}The \textit{Gaia}-ESO line lists do not fulfill this requirement.}\newline
Detailed documentation about the NLTE abundance correction mode is published in the publication database of Uppsala University\footnote{\url{https://urn.kb.se/resolve?urn=urn:nbn:se:uu:diva-571390}}.


{\color{\revcolor}
\section{Practical Applications}\label{sec:practical_applications}
}

\texttt{webSME} provides a user-friendly front-end for spectrum analysis, allowing users to derive stellar parameters from normalized spectra, with masking options for known problematic regions (e.g. ranges contaminated by telluric lines). For the normalization process we developed the open-source \texttt{Normalizer}\footnote{\url{https://github.com/astrojohannes/normalizer}} tool, which ensures that users can provide spectra in the proper format for analysis.
\texttt{webSME} has been applied to several case studies: \\

\subsection{AI Phoenicis}
AI Phoenicis is a well-known eclipsing binary, often used as a benchmark for stellar parameter determination. The system consists of an F-type dwarf (A) and a K-type subgiant (B), making it an ideal laboratory for testing potential abundance differences due to its well-determined stellar parameters and recently disentangled spectra \citep{Maxted2020}. Since both stars presumably formed from the same material, differences in their atmospheric composition may reflect evolutionary processes like atomic diffusion.

{\color{\revcolor}
As a sanity check, we inferred \(T_{\mathrm{eff}}\) and \(\log g\) in separate \texttt{webSME} runs. Previously published values \citep[e.g.][]{Andersen1988} were successfully reproduced within uncertainties. More recent reference values for AI Phe A and B are effective temperatures of \(6199 \pm 22\,\mathrm{K}\) and \(5094 \pm 16\,\mathrm{K}\), respectively, as reported by \citet{Miller2020}, and surface gravities of \(\log g = 4.002 \pm 0.001\) and \(3.598 \pm 0.001\), respectively, as derived by \citet{Maxted2020}, where \(g\) is expressed in \(\mathrm{cm\,s^{-2}}\). The agreement with the published values provides an independent validation of the parameter-inference procedure.
}

Using \texttt{webSME}, we determined the iron abundance of both components.

An essential first step is the accurate determination of the microturbulence velocity ($v_{\mathrm{mic}}$), which accounts for small-scale, non-thermal motions in the stellar photosphere that broaden spectral lines beyond thermal and pressure effects. Weak lines are mostly unaffected by $v_{\mathrm{mic}}$, because the absorbing photons mainly get re-distributed along the line shape, but the total number of absorptions remains constant. However, stronger lines with already saturated cores may become even stronger, because the re-distribution from the line core to the wings effectively adds absorbing photons. Hence, if $v_{\mathrm{mic}}$ is underestimated, the resulting abundances will systematically decrease with increasing equivalent width (EW); conversely, overestimation leads to increasing abundances with EW. Therefore, before finalizing any abundance determination, it is crucial to calibrate $v_{\mathrm{mic}}$ by removing any observable trend between abundance and EW.

\begin{figure}
    \centering
    \includegraphics[width=1\columnwidth]{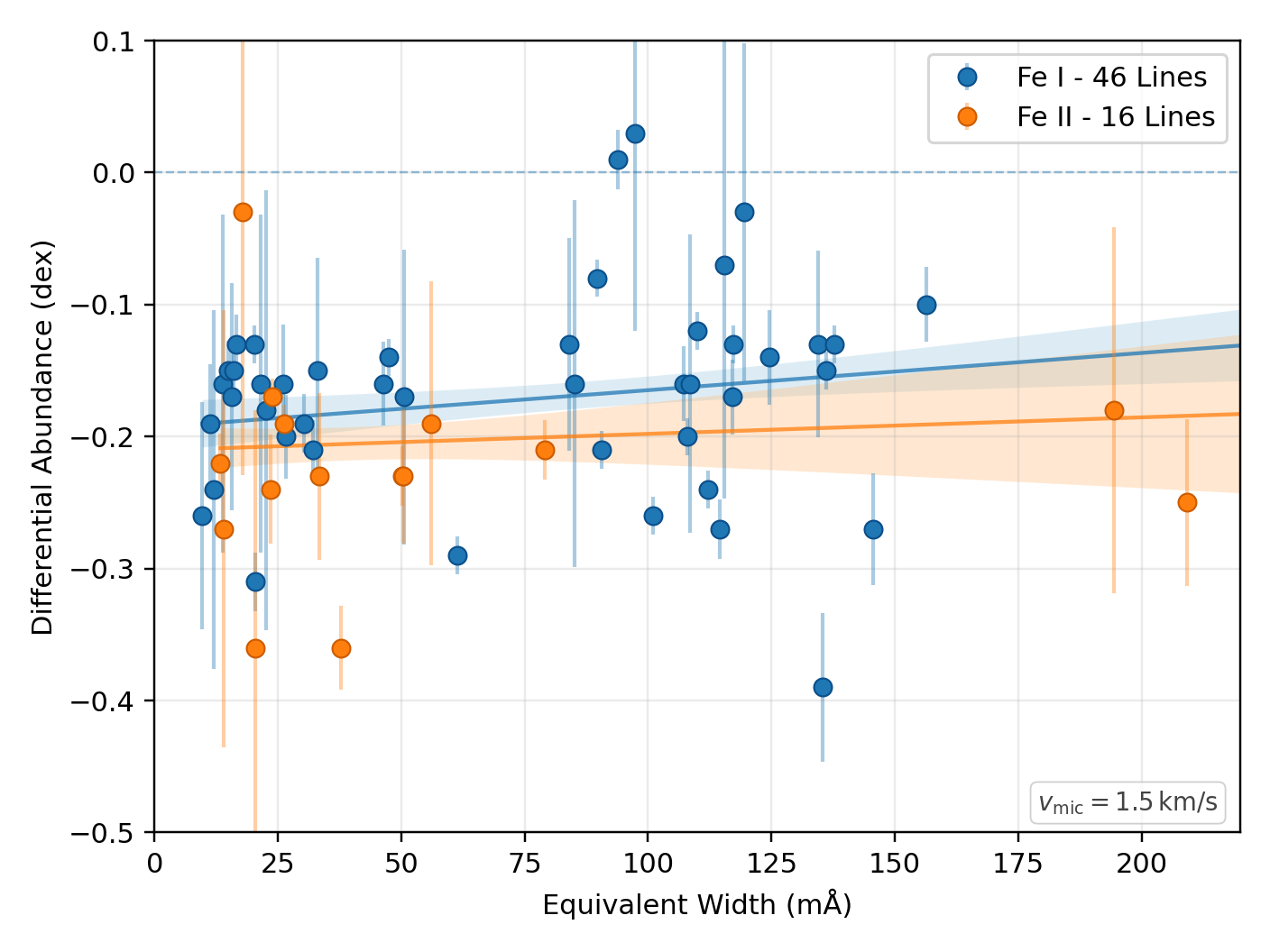}
    \caption{\ion{Fe}{i} and \ion{Fe}{ii} differential abundances vs.\ equivalent width for AI~Phe~A using \texttt{webSME}. The lack of a trend confirms the derived microturbulence velocity, $v_{\mathrm{mic}} = 1.5$~km\,s$^{-1}$.}
    \label{fig:vmic_fe_A}
\end{figure}

\begin{figure}
    \centering
    \includegraphics[width=1\columnwidth]{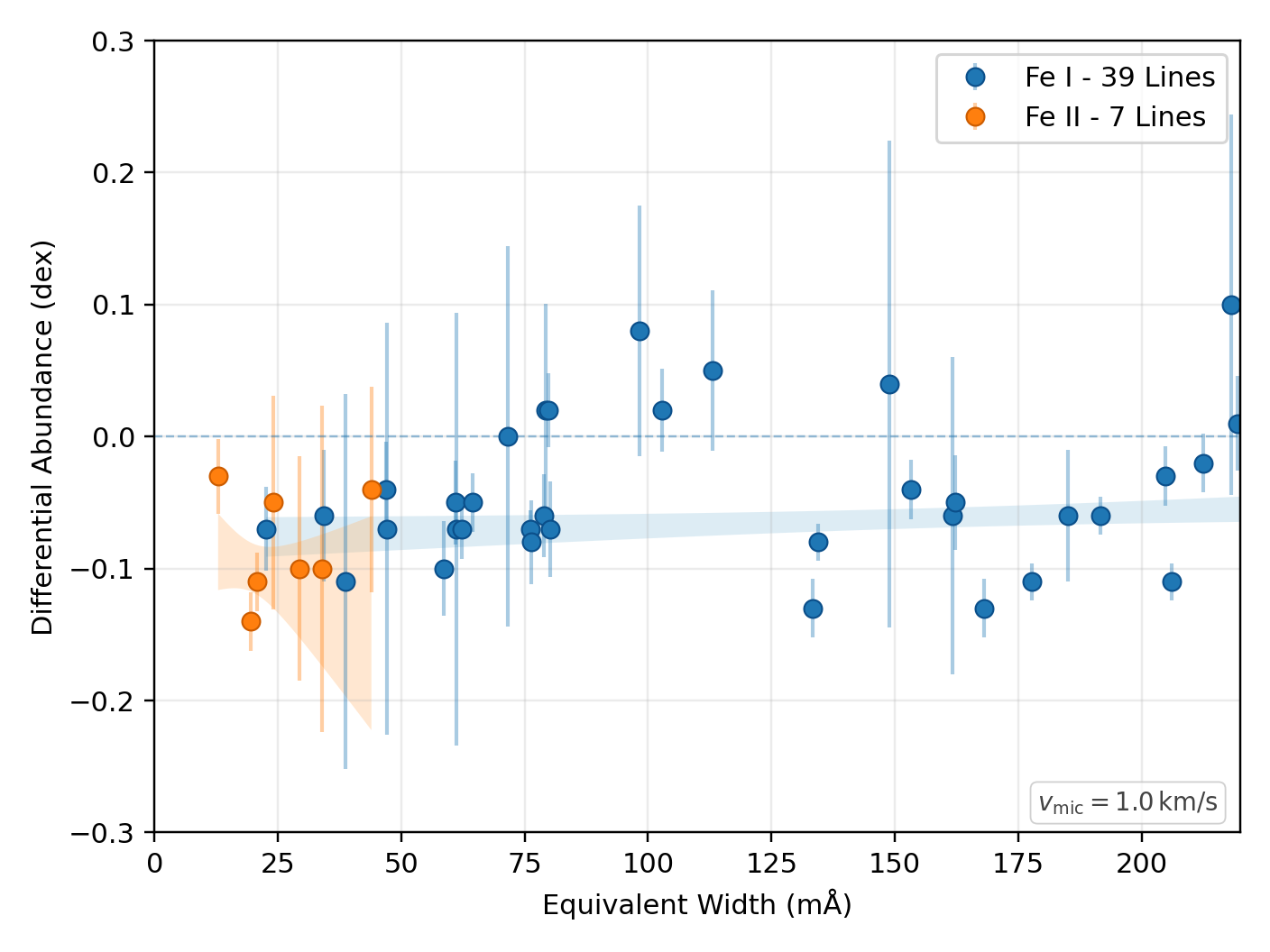}
    \caption{\ion{Fe}{i} and \ion{Fe}{ii} differential abundances vs.\ equivalent width for AI~Phe~B. No significant trend is observed, supporting $v_{\mathrm{mic}} = 1.0$~km\,s$^{-1}$.}
    \label{fig:vmic_fe_B}
\end{figure}

{\color{\revcolor}
In this study, a line-by-line fitting strategy was employed using carefully selected \ion{Fe}{i} and \ion{Fe}{ii} lines to empirically derive $v_{\mathrm{mic}}$ for AI~Phe~A and B. By iteratively adjusting $v_{\mathrm{mic}}$ and examining the slope of the differential abundance vs.\ EW relation, the optimal values were identified as $v_{\mathrm{mic}} = 1.5$~km\,s$^{-1}$ and $v_{\mathrm{mic}} = 1.0$~km\,s$^{-1}$ for the A and B component respectively. For these values trends between abundance and EW were flattened, as shown in Figures \ref{fig:vmic_fe_A} and \ref{fig:vmic_fe_B}.
}

The measured iron abundance for the primary component (A) is
{\color{\revcolor}
\([\mathrm{Fe/H}] = -0.16 \pm 0.03\), while the secondary (B) shows a slightly higher value of \([\mathrm{Fe/H}] = -0.07 \pm 0.02\).
}
This differential signature - primarily driven by \ion{Fe}{i} lines - may reflect the effects of evolutionary processes such as atomic diffusion, which are expected to alter surface compositions differently depending on the star's evolutionary phase (Maxted et al., submitted to MNRAS).

\subsection{HD 22064}
In the case of the eclipsing binary system HD22064, composed of an F dwarf and an M dwarf, the fundamental parameters of the primary component (A) have been tightly constrained in previous studies \citep{Maxted2023}, with the exception of its metallicity. In this work, we use \texttt{webSME} to constrain its metallicity. This is critical for several reasons, particularly in the context of testing and improving stellar evolution models. Metallicity is one of the fundamental parameters that significantly influence a star's internal structure, energy generation, and overall evolutionary path. In an eclipsing binary system, where both stars are assumed to have formed simultaneously from the same molecular cloud, the measurement of metallicity for each companion allows for a stringent test of this assumption. Deviations in metallicity between the two stars could indicate complex processes such as mass transfer or diffusion partially inhibited by convection or other mixing processes. Accurate metallicity measurements provide crucial input for stellar evolution models, as they affect key properties like opacity and nuclear reaction rates, particularly in low-mass stars, where even slight changes in metallicity can lead to substantial variations in predicted radii, luminosities, and lifetimes.

In the following, we present a high-resolution spectroscopic analysis of the primary component of the eclipsing binary to infer metallicity \([\mathrm{M/H}]\), elemental abundances of iron and magnesium (\(\log \epsilon(\mathrm{Fe})\), \(\log \epsilon(\mathrm{Mg})\)), and to evaluate the effect of correcting the observed spectrum for contamination from the secondary.

The analysis was based on two spectra obtained with the high-resolution FIES spectrograph. The faint M dwarf contributes approximately one percent of the total flux in the optical regime, resulting in minor spectral contamination. To address this, we generated corrected versions of the spectra by modeling the composite system using synthetic spectral templates for both stars. Flux ratios and Doppler shifts were applied based on TESS photometry and synthetic atmosphere models, enabling the extraction of a cleaned primary spectrum.

Both the corrected and uncorrected spectra were analyzed using \texttt{webSME}. As presented in the previous section, the microturbulence velocity \(v_{\mathrm{mic}}\) was determined by minimizing trends between abundance and equivalent width for a set of carefully selected iron lines. A value of \(v_{\mathrm{mic}} = 1.45\,\mathrm{km\,s^{-1}}\) was found to be optimal in both spectra. The metallicity derived from the corrected spectra was \([\mathrm{M/H}] = -0.04 \pm 0.09\), consistent with photometric estimates.

Abundances were inferred from individual unblended Fe\,\textsc{i} and Fe\,\textsc{ii} lines, yielding a final iron abundance of \(\log A(\mathrm{Fe}) = 7.45 \pm 0.14\). Magnesium abundances derived from weak, isolated Mg lines resulted in \(\log A(\mathrm{Mg}) = 7.45 \pm 0.06\). The effect of the companion star's flux contribution on these abundances was found to be negligible: differences between corrected and uncorrected spectra were typically within 0.01~-~0.02~dex.

\begin{figure*}
    \centering
    \includegraphics[width=\textwidth]{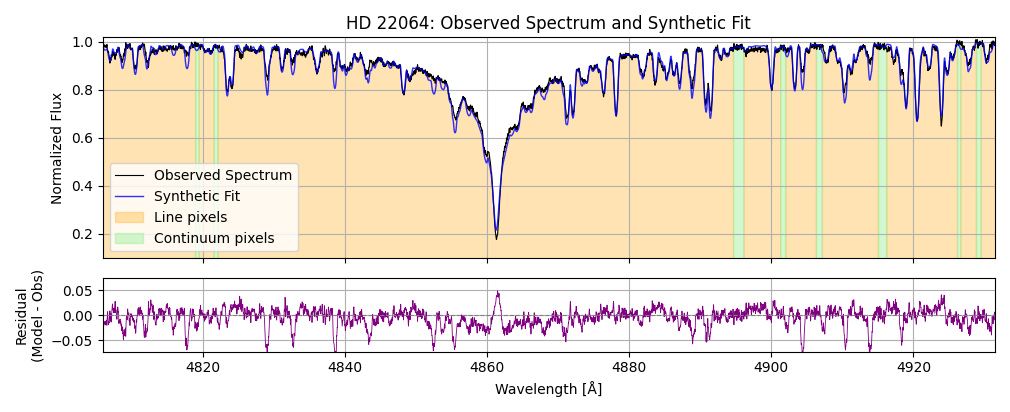}
    \caption{Normalized spectrum of HD~22064A around the H$\beta$ region (top panel), with colors indicating masked regions corresponding to lines (yellow) and continuum (green). The black and blue lines are the observed and synthetic (model) spectra. The bottom panel is the residual of (model - observation).}
    \label{fig:hd22064_hbeta}
\end{figure*}

An independent spectroscopic estimate of the effective temperature was obtained through fitting the H\,$\beta$ line (see Figure \ref{fig:hd22064_hbeta}). These results yielded a value of 6703~$\pm$~100~K, in good agreement with the literature value of \(6763 \pm 39\,\mathrm{K}\) \citep{Maxted2023}.

Hence, we conclude that HD~22064A is a near-solar metallicity F-type star. The correction for the M dwarf companion's flux has a negligible impact on the derived stellar parameters.

{\color{\revcolor}
\subsection{\texttt{webSME} in Education}\label{sec:education}
}

\texttt{webSME} can serve as an educational tool, particularly in spectroscopic training sessions like those conducted in the
ChETEC-INFRA Observational Schools 2023 and 2025, where it was used by students in conjunction with remote observations with the Nordic Optical Telescope (NOT).
Spectra from the NOT's high-resolution fibre-fed Echelle spectrograph were analyzed by the students on-site during the week to infer stellar parameters.
Several stars were observed, including HD 22064, and the tool's ability to handle diverse datasets and astrophysical sources was demonstrated.

Another example is the masterclass\footnote{\url{http://mc.chetec-infra.eu/}}, where students work with real spectra to derive stellar parameters. For instance, they analyze the H alpha and Ca absorption lines in stellar spectra to determine effective temperature, surface gravity, and metallicity. Additionally, the masterclass guides students through measuring Li abundances in metal-poor stars, which involves using weak Li lines and emphasizes the importance of accurate parameter estimation for successful analysis. This hands-on approach enhances the understanding of how spectroscopic data can be interpreted in a wider context such as the cosmological lithium problem. The masterclass teaches both the basic technical skills of working with spectral data and the scientific reasoning required to make informed adjustments during spectral fitting.

During the masterclass, it became evident that abundance derivations in some cases can fail when the metallicity-scaled reference abundance - used as the initial guess in the fitting procedure - deviates significantly from the star’s true element abundance. In such situations, it is advisable to adjust the reference abundance of the element of interest by providing a user-defined abundance pattern within \texttt{webSME} that more accurately reflects the expected composition of the star. Doing so improves the convergence of the minimization algorithm and leads to more reliable abundance determinations. By default, the starting values for the user-defined pattern are based on the solar abundances from \citet{Asplund2021}. \\


\section{Discussion and Conclusion}

{\color{\revcolor}
For robust parameter and abundance inference with \texttt{webSME} we recommend high-resolution spectra, approximately $R\gtrsim50\,000$, and \rev{S/N} $\gtrsim50$ per pixel in the analyzed regions. The current precomputed grid and line-list selections are optimized primarily for FGK stars. Cooler M-dwarf spectra require additional caution because molecular opacity and line blending become increasingly important. Telluric regions, inter-order gaps, and bad pixels should be masked. The default parameter-inference mode relies on one-dimensional model atmospheres; NLTE abundance corrections are available for the elements covered by the implemented departure-coefficient grids, but this does not constitute a full 3D NLTE treatment.
}

As demonstrated in Section \ref{sec:testing_inference}, under idealized conditions -- specifically when the input spectrum is synthetic, noise-free, and fully consistent with the line list used to compute the synthetic spectrum -- \texttt{webSME} is capable of inferring stellar parameters with high accuracy. In such cases, a ``single-click'' analysis, i.e. uploading a normalized spectrum that covers a broad wavelength range and executing LSQ parameter inference, can yield near-perfect matches with the true input parameters. This showcases the potential of \texttt{webSME} to act as a highly accessible and powerful tool in controlled environments.

However, in more realistic settings where Gaussian noise is introduced (as detailed in Section \ref{sec:testing_inference_noise}), the reliability of the inferred parameters diminishes significantly, especially at signal-to-noise ratios below 50. For instance, while the effective temperature ($T_{\mathrm{eff}}$) and metallicity ([M/H]) remain relatively stable, parameters such as surface gravity ($\log g$) and microturbulence velocity ($v_{\mathrm{mic}}$) become highly uncertain, with error bars increasing to unacceptable levels. This instability reflects the sensitivity of these parameters to degeneracies and the influence of spectral noise.

In fully realistic applications -- such as the analysis of benchmark stars HD84937 and $\beta$ Gem discussed in Section \ref{sec:testing_benchmark} -- \texttt{webSME} demonstrates that it can robustly recover $T_{\mathrm{eff}}$ and [M/H], particularly when appropriate line selection and masking (e.g. telluric regions, continuum points) are applied. Nevertheless, the inference of $\log g$ and $v_{\mathrm{mic}}$ remains a challenge due to intrinsic degeneracies in the spectral features and limited sensitivity of the line wings in noisy or blended regions. These results suggest that while the ``single-click'' approach can serve as a useful first step, it should not be solely relied upon for comprehensive stellar parameter determination.

Instead, we recommend a step-wise inference strategy: initially derive a full set of parameters from a large wavelength range using LSQ in the precomputed grid mode, then refine $T_{\mathrm{eff}}$ using Balmer lines (e.g., H$\beta$ or H$\alpha$), and [M/H] using broader ranges rich in metallic lines. Finally, $v_{\mathrm{mic}}$ should be determined through a line-by-line approach by e.g. minimizing trends between elemental abundance and line equivalent width, as demonstrated for AI Phoenicis in Section \ref{sec:practical_applications}. This layered methodology improves robustness and reflects established best practices in stellar spectroscopy.

Another noteworthy limitation arises in cases where the initial abundance of a particular element deviates significantly from the metallicity-scaled reference abundance. As discussed in Section \ref{sec:education}, this can lead to complete failure of the LSQ optimization algorithm implemented in pySME. This issue is especially relevant when analyzing stars with unusual abundance patterns, e.g. targeting elements like lithium, which can vary dramatically from solar-scaled expectations. If the initial guess is too far from the true value, the optimizer may converge poorly -- or not at all -- resulting in spurious or non-physical solutions. In such cases, it is strongly advised to provide user-defined starting abundances that more closely reflect the expected elemental composition of the star. Doing so can significantly improve convergence behavior and enhance the reliability of the inferred abundances. This again highlights the importance of informed input settings, even when using automated tools like \texttt{webSME}.


\begin{acknowledgements}
This project has received funding from the European Union's Horizon 2020 research and innovation programme under grant agreement No 101008324 (ChETEC-INFRA).\\
We thank Mingjie Jian (Stockholm U) and Henrik Jönsson (Malmö U) for their help in debugging PySME. 
\end{acknowledgements}


\bibliographystyle{aa}
\bibliography{websme}


\begin{appendix}

\section{Fe line analysis of AI Phe A }

{\color{\revcolor}
Table~\ref{tab:fe_lines_ai_phe_ab} shows the lines used for iron abundance derivations from individual \ion{Fe}{i} and \ion{Fe}{ii} species in the spectrum of AI~Phe~A and AI~Phe~B using \texttt{webSME}. Each line is identified by its central wavelength in \AA. The equivalent width (EW) of each line is given in milli-\AA~(m\AA), serving as an indicator of line strength. The columns labeled log $A_{\texttt{A}}$ and log $A_{\texttt{B}}$ list the absolute logarithmic abundances for AI Phe A and B on the usual scale where log $A(\mathrm{H}) = 12$. Solar differential abundances are found in columns denoted by [Fe/H]$_{\mathrm{A}}$ and [Fe/H]$_{\mathrm{B}}$ for AI~Phe~A and AI~Phe~B respectively.

All individual line fits for AI Phe A and the corresponding solar lines are available at
\url{https://www.quellcode.gmbh/websme/paper/websme_tasks_summary_AIPheA_run8.html}
and \url{https://www.quellcode.gmbh/websme/paper/websme_tasks_summary_HARPS_Sun_normalized_run8.html},
respectively. Likewise, the line fits for AI Phe B and the corresponding solar lines are available at
\url{https://www.quellcode.gmbh/websme/paper/websme_tasks_summary_AIPheB_run1.html}
and \url{https://www.quellcode.gmbh/websme/paper/websme_tasks_summary_HARPS_Sun_normalized_run1.html},
respectively.

The line selection was curated to include preferably isolated and only weakly blended features with a reliable local continuum placement. This is important because both abundance and EW measurements are sensitive to continuum-normalisation errors and contamination from neighbouring lines. Lines affected by severe blending, poor continuum definition, or visibly problematic profile fits were therefore avoided where possible.
}

{\color{\revcolor}
For the \texttt{webSME} analysis, the abundances were inferred by spectral synthesis using the atomic GaiaESO line list. The stellar parameters were kept fixed. The EWs listed in the table were not obtained by a direct integration of the observed profile. Instead, they were derived by forward modeling. After the best-fitting synthetic line profile had been obtained using fixed stellar parameters and the corresponding line abundance, the EW was measured by forward modeling  of each single line of interest. This procedure ensures that the EW refers to the same physical line model that was used for the abundance determination and reduces the impact of residual blends or noise spikes in the observed spectrum.
}


\begin{table*}[t]
\centering
{\color{\revcolor}
\caption{Fe-line analysis for AI Phe A and AI Phe B. The EW values are only reported as an indication of the line strengths, they are not used in the derivation of the abundances.}
\label{tab:fe_lines_ai_phe_ab}
\tiny
\setlength{\tabcolsep}{0pt}
\renewcommand{\arraystretch}{0.96}

\begin{tabular*}{\textwidth}{@{\extracolsep{\fill}}llrrrrrr@{}}
\toprule
& & \multicolumn{3}{c}{AI Phe A} & \multicolumn{3}{c}{AI Phe B} \\
\cmidrule(lr){3-5}\cmidrule(lr){6-8}
$\lambda_c$ (\AA) & Species &
EW (m\AA) & log $A_{\mathrm{A}}$ & [Fe/H]$_{\mathrm{A}}$ &
EW (m\AA) & log $A_{\mathrm{B}}$ & [Fe/H]$_{\mathrm{B}}$ \\
\midrule
	4787.8266 & Fe I & $21$ & $7.30\pm0.08$ & $-0.160\pm0.128$ & $61$ & $7.39\pm0.13$ & $-0.070\pm0.164$ \\
	4807.7082 & Fe I & $33$ & $7.44\pm0.03$ & $-0.150\pm0.085$ & $72$ & $7.59\pm0.12$ & $0.000\pm0.144$ \\
	4808.1478 & Fe I & $14$ & $7.39\pm0.08$ & $-0.160\pm0.128$ & $47$ & $7.48\pm0.12$ & $-0.070\pm0.156$ \\
	4893.8126 & Fe II & $20$ & $7.33\pm0.06$ & $-0.360\pm0.180$ & $--$ & $--$ & $--$ \\
	4920.5023 & Fe I & $235$ & $7.28\pm0.01$ & $-0.080\pm0.014$ & $--$ & $--$ & $--$ \\
	4923.9212 & Fe II & $194$ & $7.28\pm0.07$ & $-0.180\pm0.139$ & $--$ & $--$ & $--$ \\
	4977.6481 & Fe I & $16$ & $7.38\pm0.05$ & $-0.170\pm0.086$ & $--$ & $--$ & $--$ \\
	5018.4356 & Fe II & $209$ & $7.21\pm0.02$ & $-0.250\pm0.063$ & $--$ & $--$ & $--$ \\
	5049.8197 & Fe I & $114$ & $7.28\pm0.02$ & $-0.270\pm0.022$ & $260$ & $7.64\pm0.02$ & $0.090\pm0.022$ \\
	5054.6425 & Fe I & $20$ & $7.16\pm0.01$ & $-0.310\pm0.022$ & $59$ & $7.37\pm0.03$ & $-0.100\pm0.036$ \\
	5133.6882 & Fe I & $125$ & $7.13\pm0.02$ & $-0.140\pm0.036$ & $185$ & $7.21\pm0.04$ & $-0.060\pm0.050$ \\
	5169.0282 & Fe II & $302$ & $7.55\pm0.02$ & $-0.030\pm0.073$ & $--$ & $--$ & $--$ \\
	5194.9414 & Fe I & $115$ & $7.34\pm0.13$ & $-0.070\pm0.177$ & $218$ & $7.51\pm0.08$ & $0.100\pm0.144$ \\
	5198.7108 & Fe I & $84$ & $7.38\pm0.04$ & $-0.130\pm0.081$ & $--$ & $--$ & $--$ \\
	5225.5261 & Fe I & $51$ & $7.39\pm0.05$ & $-0.170\pm0.112$ & $--$ & $--$ & $--$ \\
	5250.6456 & Fe I & $85$ & $7.42\pm0.07$ & $-0.160\pm0.139$ & $--$ & $--$ & $--$ \\
	5256.9319 & Fe II & $24$ & $7.37\pm0.01$ & $-0.240\pm0.041$ & $24$ & $7.56\pm0.07$ & $-0.050\pm0.081$ \\
	5264.8024 & Fe II & $56$ & $7.30\pm0.04$ & $-0.190\pm0.108$ & $--$ & $--$ & $--$ \\
	5339.9293 & Fe I & $117$ & $7.30\pm0.02$ & $-0.170\pm0.028$ & $219$ & $7.48\pm0.03$ & $0.010\pm0.036$ \\
	5364.8709 & Fe I & $109$ & $7.28\pm0.08$ & $-0.160\pm0.113$ & $162$ & $7.38\pm0.09$ & $-0.060\pm0.120$ \\
	5373.7086 & Fe I & $48$ & $7.24\pm0.01$ & $-0.140\pm0.014$ & $76$ & $7.31\pm0.01$ & $-0.070\pm0.014$ \\
	5383.3685 & Fe I & $138$ & $7.16\pm0.01$ & $-0.130\pm0.014$ & $206$ & $7.18\pm0.01$ & $-0.110\pm0.014$ \\
	5393.1672 & Fe I & $108$ & $7.23\pm0.01$ & $-0.200\pm0.014$ & $192$ & $7.37\pm0.01$ & $-0.060\pm0.014$ \\
	5397.1279 & Fe I & $156$ & $7.37\pm0.02$ & $-0.100\pm0.028$ & $378$ & $7.44\pm0.02$ & $-0.030\pm0.028$ \\
	5410.9097 & Fe I & $107$ & $7.13\pm0.02$ & $-0.160\pm0.028$ & $162$ & $7.24\pm0.03$ & $-0.050\pm0.036$ \\
	5414.0698 & Fe II & $33$ & $7.28\pm0.02$ & $-0.230\pm0.063$ & $29$ & $7.41\pm0.06$ & $-0.100\pm0.085$ \\
	5425.2485 & Fe II & $50$ & $7.24\pm0.01$ & $-0.230\pm0.051$ & $44$ & $7.43\pm0.06$ & $-0.040\pm0.078$ \\
	5434.5235 & Fe I & $136$ & $7.29\pm0.01$ & $-0.150\pm0.014$ & $297$ & $7.40\pm0.01$ & $-0.040\pm0.014$ \\
	5491.8315 & Fe I & $--$ & $--$ & $--$ & $23$ & $7.36\pm0.03$ & $-0.070\pm0.032$ \\
	5522.4461 & Fe I & $--$ & $--$ & $--$ & $61$ & $7.50\pm0.03$ & $-0.050\pm0.032$ \\
	5525.1170 & Fe II & $14$ & $7.19\pm0.07$ & $-0.270\pm0.166$ & $--$ & $--$ & $--$ \\
	5546.9905 & Fe I & $12$ & $7.25\pm0.08$ & $-0.240\pm0.136$ & $39$ & $7.38\pm0.09$ & $-0.110\pm0.142$ \\
	5569.6180 & Fe I & $112$ & $7.21\pm0.01$ & $-0.240\pm0.014$ & $213$ & $7.43\pm0.02$ & $-0.020\pm0.022$ \\
	5572.8423 & Fe I & $146$ & $7.37\pm0.03$ & $-0.270\pm0.042$ & $--$ & $--$ & $--$ \\
	5705.4642 & Fe I & $23$ & $7.21\pm0.09$ & $-0.180\pm0.166$ & $--$ & $--$ & $--$ \\
	5855.0758 & Fe I & $11$ & $7.23\pm0.02$ & $-0.190\pm0.045$ & $34$ & $7.36\pm0.03$ & $-0.060\pm0.050$ \\
	5916.2473 & Fe I & $32$ & $7.38\pm0.01$ & $-0.210\pm0.022$ & $80$ & $7.52\pm0.03$ & $-0.070\pm0.036$ \\
	5991.3709 & Fe II & $38$ & $7.36\pm0.01$ & $-0.360\pm0.032$ & $34$ & $7.62\pm0.12$ & $-0.100\pm0.124$ \\
	6082.7101 & Fe I & $16$ & $7.34\pm0.01$ & $-0.150\pm0.014$ & $62$ & $7.42\pm0.02$ & $-0.070\pm0.022$ \\
	6084.1017 & Fe II & $26$ & $7.38\pm0.01$ & $-0.190\pm0.014$ & $21$ & $7.46\pm0.02$ & $-0.110\pm0.022$ \\
	6094.3728 & Fe I & $10$ & $7.53\pm0.05$ & $-0.260\pm0.086$ & $--$ & $--$ & $--$ \\
	6113.3192 & Fe II & $13$ & $7.37\pm0.01$ & $-0.220\pm0.022$ & $13$ & $7.56\pm0.02$ & $-0.030\pm0.028$ \\
	6136.6149 & Fe I & $119$ & $7.43\pm0.08$ & $-0.030\pm0.128$ & $--$ & $--$ & $--$ \\
	6151.6173 & Fe I & $27$ & $7.30\pm0.01$ & $-0.200\pm0.032$ & $79$ & $7.44\pm0.01$ & $-0.060\pm0.032$ \\
	6200.3125 & Fe I & $--$ & $--$ & $--$ & $103$ & $7.57\pm0.03$ & $0.020\pm0.032$ \\
	6226.7342 & Fe I & $15$ & $7.35\pm0.01$ & $-0.150\pm0.022$ & $47$ & $7.46\pm0.03$ & $-0.040\pm0.036$ \\
	6229.2259 & Fe I & $20$ & $7.26\pm0.01$ & $-0.130\pm0.014$ & $64$ & $7.34\pm0.02$ & $-0.050\pm0.022$ \\
	6230.7222 & Fe I & $117$ & $7.35\pm0.01$ & $-0.130\pm0.014$ & $205$ & $7.45\pm0.02$ & $-0.030\pm0.022$ \\
	6232.6403 & Fe I & $61$ & $7.27\pm0.01$ & $-0.290\pm0.014$ & $113$ & $7.61\pm0.06$ & $0.050\pm0.061$ \\
	6239.9427 & Fe II & $18$ & $7.46\pm0.06$ & $-0.030\pm0.199$ & $--$ & $--$ & $--$ \\
	6240.6462 & Fe I & $26$ & $7.27\pm0.02$ & $-0.160\pm0.045$ & $79$ & $7.45\pm0.07$ & $0.020\pm0.081$ \\
	6246.3180 & Fe I & $91$ & $7.24\pm0.01$ & $-0.210\pm0.014$ & $153$ & $7.41\pm0.02$ & $-0.040\pm0.022$ \\
	6270.2234 & Fe I & $30$ & $7.18\pm0.01$ & $-0.190\pm0.022$ & $80$ & $7.39\pm0.02$ & $0.020\pm0.028$ \\
	6369.4590 & Fe II & $24$ & $7.26\pm0.01$ & $-0.170\pm0.014$ & $19$ & $7.29\pm0.02$ & $-0.140\pm0.022$ \\
	6393.6004 & Fe I & $110$ & $7.27\pm0.01$ & $-0.120\pm0.014$ & $178$ & $7.28\pm0.01$ & $-0.110\pm0.014$ \\
	6400.0003 & Fe I & $135$ & $7.35\pm0.04$ & $-0.390\pm0.057$ & $--$ & $--$ & $--$ \\
	6411.6480 & Fe I & $101$ & $7.26\pm0.01$ & $-0.260\pm0.014$ & $168$ & $7.39\pm0.02$ & $-0.130\pm0.022$ \\
	6421.3499 & Fe I & $90$ & $7.32\pm0.01$ & $-0.080\pm0.014$ & $133$ & $7.27\pm0.02$ & $-0.130\pm0.022$ \\
	6432.6757 & Fe II & $50$ & $7.26\pm0.01$ & $-0.230\pm0.022$ & $--$ & $--$ & $--$ \\
	6456.3796 & Fe II & $79$ & $7.37\pm0.01$ & $-0.210\pm0.022$ & $--$ & $--$ & $--$ \\
	6494.9804 & Fe I & $134$ & $7.43\pm0.05$ & $-0.130\pm0.071$ & $241$ & $7.47\pm0.05$ & $-0.090\pm0.071$ \\
	6498.9383 & Fe I & $17$ & $7.28\pm0.02$ & $-0.130\pm0.022$ & $76$ & $7.33\pm0.03$ & $-0.080\pm0.032$ \\
	6546.2381 & Fe I & $94$ & $7.35\pm0.02$ & $0.010\pm0.022$ & $135$ & $7.26\pm0.01$ & $-0.080\pm0.014$ \\
	6592.9124 & Fe I & $97$ & $7.32\pm0.09$ & $0.030\pm0.150$ & $149$ & $7.33\pm0.14$ & $0.040\pm0.184$ \\
	6609.1097 & Fe I & $46$ & $7.41\pm0.01$ & $-0.160\pm0.032$ & $98$ & $7.65\pm0.09$ & $0.080\pm0.095$ \\
\bottomrule
\end{tabular*}
}
\end{table*}

\vspace{3cm} 

\section{NLTE grids and model atoms}
As mentioned in Section~\ref{sec: nlte abundance corr mode}, the grids of departure coefficients used by \texttt{webSME} have all been published relatively recently, most within the past 5 years. The model atoms used to calculate these grids are also modern, most of which dating back roughly 6-8 years. A summary of references for all grids and model atoms is given in Table~\ref{tab:NLTE atoms + grids}. 


\begin{table}[hb]
    \centering
    \caption{References for grids and underlying model atoms used in NLTE calculations by \texttt{webSME}.}
    \begin{tabular}{lrr}
    \toprule
    Element & Ref. Atom & Ref. Grid\\
    \midrule
       H    & \cite{Amarsi_2018}        &   \cite{Amarsi_2020}\\
       Li   & \cite{Lind2013}           &   \cite{Amarsi_2020}\\
       C    & \cite{Amarsi2019}         &   \cite{Amarsi_2020}\\
       N    & \cite{Amarsi2020a}        &   \cite{Amarsi_2020}\\
       O    & \cite{Amarsi2018a}        &   \cite{Amarsi_2020}\\
       Na   & \cite{Lind2011}           &   \cite{Amarsi_2020}\\
       Mg   & \cite{Osorio2015}         &   \cite{Amarsi_2020}\\
       Al   & \cite{Nordlander2017}     &   \cite{Amarsi_2020}\\
       Si   & \cite{Amarsi2017}         &   \cite{Amarsi_2020}\\
       K    & \cite{Reggiani2019}       &   \cite{Amarsi_2020}\\
       Ca   & \cite{Osorio2019}         &   \cite{Amarsi_2020}\\
       Ti   & \cite{Mallinson2022}      &   \cite{Mallinson_2024}\\
       Mn   & \cite{Bergemann2019}      &   \cite{Amarsi_2020}\\
       Fe   & \cite{Lind2017}           &   \cite{Amarsi_2022}\\ 
       Cu   & \cite{Racca2025}      &   \cite{Caliskan_2025}\\ 
       Ba   & \cite{Gallagher2020}      &   \cite{Amarsi_2020}\\  
    \bottomrule
    \end{tabular}
    \label{tab:NLTE atoms + grids}
\end{table}

\end{appendix}

\end{document}